\documentclass[traditabstract]{aa}  

\usepackage{graphicx}
\usepackage{rotating}
\usepackage{txfonts}

\usepackage{amssymb}
\usepackage[fleqn]{amsmath}

\usepackage{url}

\usepackage[pdfencoding=auto,psdextra]{hyperref}
\hypersetup{
    colorlinks=true,
    linkcolor=blue,
    filecolor=magenta,      
    urlcolor=blue,
    citecolor=blue
}
\usepackage{xcolor}

\begin{document}

   \title{Self-similarity of the mass distribution in rich galaxy clusters up $z\sim1$ tracked with weak lensing}


   \author{
    Mauro Sereno \inst{\ref{inaf-oas},\ref{infn-bo}\thanks{\email{mauro.sereno@inaf.it}}}
          }              
\institute{
INAF - Osservatorio di Astrofisica e Scienza dello Spazio di Bologna, via Piero Gobetti 93/3, I-40129 Bologna, Italy \label{inaf-oas}
\and
INFN, Sezione di Bologna, viale Berti Pichat 6/2, I-40127 Bologna, Italy \label{infn-bo}
}


 \abstract{
In the standard theory of growth of the nonbaryonic dark matter, cosmic structures form hierarchically and self-similarly from smaller clumps. The assembly merger tree goes from the linear perturbations in the early universe to highly non linear structures at late times. Gravity is the driving force and self-similarity should inform cosmic haloes.
However, it is unclear if apparent anomalies at non-linear scales are due to either baryonic or new physics.
Here, I show that the mass distribution of rich haloes evolve self-similarly at least since the universe was $5.7~\text{Gyr}$ old. Using gravitational weak lensing, I constrain the mass profiles of galaxy clusters with $M_\text{200c} \gtrsim 2 \times 10^{14}M_\odot$ that were optically detected in the HSC-SSP survey in the redshift range $0.2 \leq z < 1.0$. 
Cluster self-similarity confirms the standard theory of growth in the non-linear regime. Clusters are still growing but neither violent mergers nor matter slowly falling in from the cosmic web disrupt self-similarity, which is in place well before the halo formation time. 
Dark matter growth can fit the fossil cosmic microwave background as well as young, very massive haloes. Looking with next generation surveys at scales in clusters where self-similarity breaks could pose a new challenge to dark matter.
}

    \authorrunning{M. Sereno}
    \titlerunning{Self-similarity of galaxy clusters}
    
    \maketitle
%

\section{Introduction}

In the standard $\Lambda$CDM (Cold Dark Matter with a cosmological constant $\Lambda$) cosmology, the cosmic structure grew by gravity out of departures of the primeval mass distribution from homogeneity \citep{pee25,efs25,mad25}. The agreement between linear perturbation theory and measurements of thermal cosmic microwave background (CMB) radiation at redshift $z\sim 1000$ offers a substantial piece of evidence for the model. CMB lensing, caused mainly by matter at $z \sim 2$, is consistent with the value inferred from the primary anisotropies. In contrast, late-time evolution and nonlinear scales probed by weak lensing measurements from cosmic shear surveys show some tension that could be reconciled if yet-to-be-understood processes suppress the amplitude of the nonlinear spectrum on small scales. Currently, it is unclear whether the discrepancies are due to observational systematics, baryonic effects that are not adequately explained, or new physics \citep{efs25,mad25}.

Galaxy clusters stand at the other end of the growth spectrum \citep{voi05}. CMB is a fossil, clusters are young and still growing. As the latest and most massive nearly virialised haloes to form, their properties put  important constraints on cosmological formation theories. 

Galaxy clusters are unique laboratories to probe DM and provided the first evidence for it \citep{zwi37}. Validation of the $\Lambda$CDM model with abundance of galaxy clusters would be another strong check in the non-linear regime. Some results from cluster count analyses \citep{planck_2013_XX, planck_2015_XXIV, des_cos+al21} were in tension with multiple cosmological probes, for example, supernovae, baryon acoustic oscillations, cosmic shear, galaxy clustering, or CMB anisotropies. However, other cluster analyses did not support previous results \citep{les+al22,spt_boc+24,erosita_ghi+al24}. More than signs of physics beyond the $\Lambda$CDM model, the tension might be due to unaccounted for systematic uncertainties related to incomplete or impure samples or biased mass calibration. 

Cosmological information is encoded in the mass distribution of galaxy clusters. In the standard scenario, structure grows hierarchically.  The massive DM haloes that host massive clusters have assembled through a  process that depends on the cosmic matter content, expansion rate, and amplitude of initial matter density fluctuations \citep{cor+al22}. First tests based on sparsity, that is, the ratio of the halo mass within radii enclosing different over-densities, are nearly insensitive to selection or mass calibration biases and provide constraints in agreement with primary CMB analyses \citep{cor+al21}.

The density fluctuations that give rise to massive clusters are approximately scale-free and result in a self-similar evolution apart from dissipative processes \citep{kai86}. The formation of bound, virialised structures in an expanding universe can be described as the spherical collapse of positive density perturbations \citep{gu+go72}. Bound shells continue to turn around and fall in. The secondary infall approaches self-similaritiy and informs the halo mass distribution \citep{gun77,fi+go84,ber85}.

However, the collapse of peaks is generally expected to be non spherical and to follow the preferential direction of the filamentary structure of the cosmic web. Numerical simulations of DM haloes can recover the merger tree of cosmic structure formation, as clumps of matter merge to form larger clumps in a hierarchy of mergers that continued to the formation of galaxies and groups and clusters of galaxies as galaxy merging slowed. Simulations have shown that the growth of DM halo density profiles undergoes two major phases \citep{nfw97,bul+al01,zha+al03,wec+al02,di+kr14,fuj+al18a}. In the early stages, the accretion is fast and mass builds up in the central region of the cluster. In a later, slow accretion phase, the mass builds in the outer region while the mass density in the core remains approximately constant \citep{go+re75,gun77,lo+ho00,asc+al04,asc+al07,li+da11}.

The numerical study of galaxy clusters at high redshift is a computationally demanding task, as simulations have to resolve the inner regions of the clusters down to small scales and, at the same time, they need to correctly reproduce the influence of the large-scale structure on the outer, gravity-dominated regions. First results confirm self-similarity \citep{leb+al18,mos+al19,sin+al25}. The evolution of the density profiles of the 25 most massive galaxy clusters in a DM universe, once scaled to the critical density, exhibits low dispersion and little evolution up to redshifts $z\sim 1$ \citep{leb+al18}, suggesting that self-similarity was established early in the formation process. This is confirmed by tracking the merging history of the 300 most massive clusters at $z=0$ backwards in time  \citep{mos+al19}. The mass distribution of the progenitors is already in place by $z \sim 2.5$.

The complex behaviour of baryons can hamper self-similarity in galaxies or small groups. Being DM dominated with a baryonic fraction in line with the universal value, massive galaxy clusters offer a useful laboratory for DM tests. As highly non linear structures, they might be difficult to model, but in the new era of precision astronomy they can be investigated with many tools. 

The main baryonic components consists of a hot, X-ray emitting intracluster medium (ICM). The bulk of the ICM outside the inner core  in massive clusters has evolved self-similarly since $z\sim 1.9$, as tracked by X-ray inferred gas density profiles and thermodynamic properties \citep{mcd+al17,ghi+al21}. This hints at the self-similar evolution of the underlying DM distribution, since rich clusters are DM dominated and the gas distribution, except in the very central regions, is driven by gravity.

Here, I directly measure the self-similarity of the matter profiles of rich clusters by detection of the weak lensing (WL) signal. The theory of WL by galaxy clusters is well understood \citep{ba+sc01,ume20}. WL distorts the shape of the background galaxies and the shear profile accurately and precisely tracks the total mass distribution of the lens \citep{hoe+al12, wtg_I_14, wtg_III_14, hoe+al15, ume+al14, ume+al16b,ok+sm16,die+al19}. WL provides a direct measurement of the excess surface mass density of the cluster acting as a lens, with no need for assumptions about the dynamical status or the equilibrium between baryons and DM. 

Until relatively recently, WL analyses were only possible with dedicated, targeted observations \citep{wtg_I_14,ume+al14}, which made the study of statistically complete, homogeneous, and large samples elusive. Photometric surveys have now advanced to so-called Stage-III \citep{alb+al06}. Thanks to their ever increasing depth, the WL signal of clusters samples can now be measured up to high redshifts \citep{ser+al17_psz2lens,ume+al20,mel+al15,hsc_med+al18b,ser+al18_psz2lens}. \citet{sin+al25} measured the WL signal of galaxy clusters in the area covered by the Dark Energy Survey (DES) which were first selected by the South Pole Telescope thermal Sunyaev-Zel’dovich effect and then optically confirmed. They found some evidence for self-similarity up to $z\sim 0.6$. Here, I take advantage of the deepest Stage-III galaxy image survey to look for self-similarity up to $z\sim 1.0$.

As reference cosmological framework, I assume the concordance flat $\Lambda$CDM model with total matter density parameter $\Omega_\text{M}=0.3$, baryonic parameter $\Omega_\text{B}=0.05$, Hubble constant $H_0=70~\text{km~s}^{-1}\text{Mpc}^{-1}$, power spectrum amplitude $\sigma_8=0.8$ and initial index $n_\text{s}=1$. When $H_0$ is not specified, $h$ is the Hubble constant in units of $100~\text{km~s}^{-1}\text{Mpc}^{-1}$. The suffix $\Delta\text{c}$ refers to the region within which the halo density is $\Delta$ times the cosmological critical density at the cluster redshift, $\rho_\text{cr}$. `$\log$' is the logarithm in base 10, and `$\ln$' is the natural logarithm.


\section{The deepest Stage-III survey}
\label{sec_HSC-SSP}

\begin{figure}
\resizebox{\hsize}{!}{\includegraphics{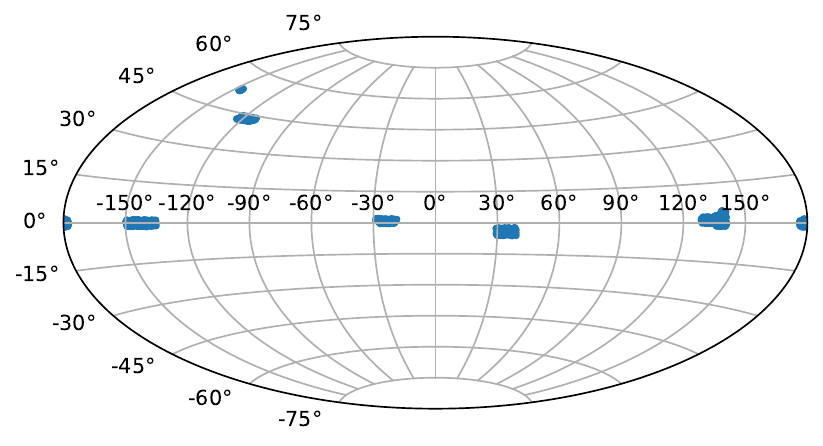}} 
\caption{Sky coverage in equatorial coordinates of the lensing survey HSC-SSP S16a used in this work.}
\label{fig_survey}
\end{figure}

The Hyper Suprime-Cam Subaru Strategic Program \citep[HSC-SSP,][]{hsc_miy+al18,hsc_aih+al18} is the deepest Stage-III galaxy image survey and it opens a window on cluster physics at high redshift.
 HSC-SSP is a multi-band imaging survey in five optical bands ($g$, $r$, $i$, $z$, $y$) with HSC, an optical wide-field imager with a field-of-view of $1.77\deg^2$  mounted on the prime focus of the $8.2\,\text{m}$ Subaru telescope \citep{hsc_miy+al18,hsc_kom+al18,hsc_fur+al18,hsc_kaw+al18}. The wide survey is optimised for WL  studies \citep{hsc_man+al18,hsc_hik+al19,hsc_miy+al19,hsc_ham+al20}, and it aims to observe $\sim1200\,\deg^2$ with a depth of $i \sim26~\text{mag}$ at the $5\,\sigma$ limit within a 2 arcseconds diameter aperture \citep{hsc_aih+al18}. 
 
Here, I consider the latest data releases at the time of writing for each product used in the analysis, which are S16a for the shear catalogue, see Sec.~\ref{sec_shea}, and S18a for the catalogue of clusters, see Sec.~\ref{sec_clus}. The analysis is limited to the clusters in the S16a footprint.

 
 \subsection{Shears}
 \label{sec_shea}
 
Here, I exploit, the S16a catalogue of galaxy shape measurements from the first-year data release \citep{hsc_man+al18}. The catalogue covers an area of $\sim 137\deg^2$ split into six fields, see Fig.~\ref{fig_survey}, observed to final depth, with a mean $i$-band seeing of 0.58 arcseconds. Galaxy shapes were estimated on the co-added $i$-band images \citep{hi+se03}, fitting a Gaussian profile with elliptical isophotes to the image with a conservative magnitude cut of $i < 24.5~\text{mag}$. The nominal unweighted and weighted source number densities are of 24.6 and 21.8 arcmin$^{-2}$, respectively \citep{hsc_man+al18}.

My choices for the lensing analysis are informed from previous WL studies \citep{hsc_chi+al20,ume+al20}. Photometric redshifts (photo-$z$  or $z_\text{p}$) are well  recovered in the redshift range $0.2 \lesssim z \lesssim 1.5$, with an accuracy of $\sigma_{z_\text{p}} / (1 + z_\text{p}) \sim 0.05~(0.04)$, and an outlier rate of $\sim15~(8)\,\%$ for galaxies down to $i$ = 25~(24) \citep{hsc_tan+al18}. I consider photo-$z$s measured with the \texttt{EPHOR\_AB} code based on PSF-matched aperture photometry \citep{hsc_tan+al18}. For the photometric magnitudes, I consider the forced \texttt{cmodel}  \citep{hsc_hua+al18}.


\subsection{Clusters}
\label{sec_clus}

\begin{figure}
\resizebox{\hsize}{!}{\includegraphics{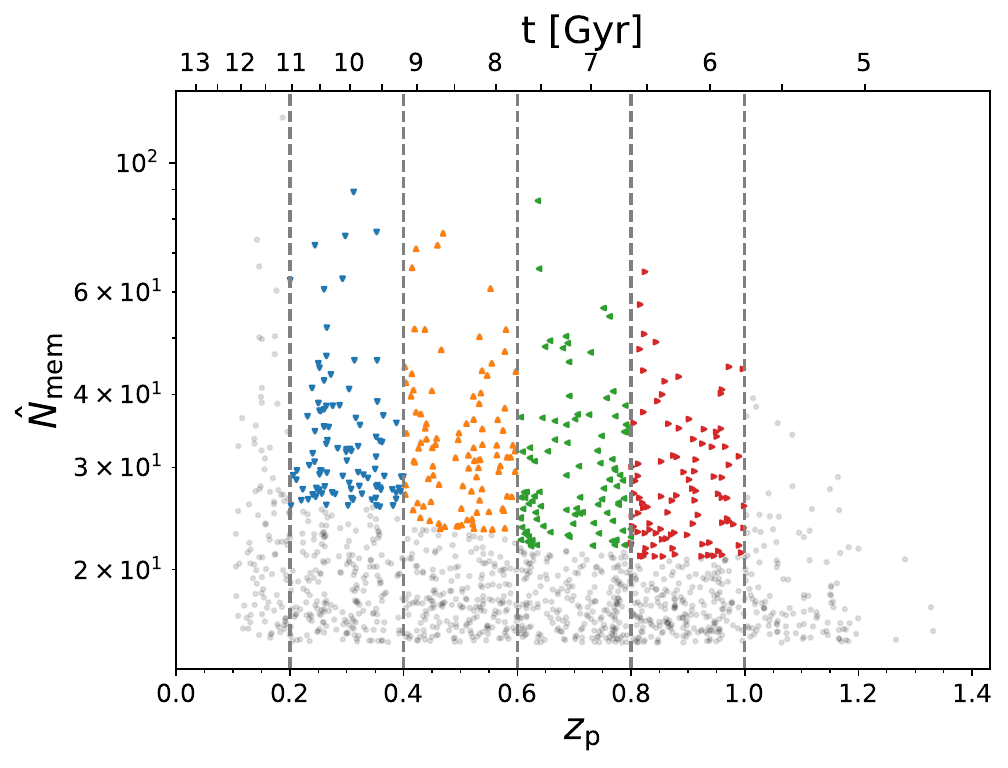}} 
\caption{
Distribution of the CAMIRA clusters in the redshift - richness plane. The age of the Universe at a given redshift is plotted on the top axis. Points are colour coded according to their redshift bin. Clusters represented by grey points are excluded from the WL analysis.}
\label{fig_z_N_mem}
\end{figure}

A complete and homogeneous sample of optically-selected clusters was retrieved from HSC-SSP with the Cluster finding algorithm based on Multi-band Identification of Red sequence gAlaxies \citep[CAMIRA,][]{ogu+al18}. CAMIRA fits each galaxy in the image with a stellar population synthesis model to compute the likelihood to be a red sequence galaxy at a given redshift \citep{ogu14,ogu+al18}. Each cluster candidate is assigned a redshift and a richness $\hat{N}_\text{mem}$ which describes the number of red member galaxies with stellar masses $M_{*}  \gtrsim 10^{10.2} M_\odot$ and within a circular aperture with a radius $R \lesssim 1.4~\text{Mpc}$. The richness is a reliable mass proxy, with an intrinsic scatter of $\sim 15\,\%$ at fixed mass \citep{ogu+al18}.

Here, I consider the catalogue version constructed from the public data release 3 (S18a). There are 1439 clusters in the WL footprint at redshift $0.1<z<1.1$ and with richness $\hat{N}_\text{mem}>15$, that roughly corresponds to $M_\text{200c} \gtrsim 1.1 \times 10^{14} M_\odot$ \citep{ogu+al18}. The sample shows high purity ($\gtrsim 95\%$) down to the richness limit of $\hat{N}_\text{mem}=15$ and high completeness ($\sim95\%$) for high mass clusters $M_\text{200c} \gtrsim 2 \times 10^{14} M_\odot$. The cluster redshift is recovered with a scatter of $0.008 \times (1+z)$. Nearly 68\,\% of the clusters is well centred, the others show a mean off-set of $\sim0.4~\text{Mpc}$. 

I study the evolution of clusters density profiles in four equally spaced redshifts bins from $z=0.2$ to 1, see Fig.~\ref{fig_z_N_mem}. In each bin, I select the 100 richest clusters to focus on the high mass end of the halo mass function. Clusters at $z<0.2$ are excluded from the analysis due to the small comoving volume. Clusters at $z\geq1$ are excluded due to the poor WL signal.

\section{Methods}
\label{sec_meth}


Here, I present the methods used for WL inference of mass and shape profiles, for abundance matching, and for profile comparison.

\subsection{Signal definition}

The main observable to constrain density profiles with lensing is the reduced excess surface density $\Delta \Sigma_{g{\rm t}}$, which can be expressed in terms of the surface density, $\Sigma$, and of the excess surface density $\Delta \Sigma_\text{t}$. For an axially symmetric lens and a single source plane, 

\begin{equation}
\Delta \Sigma_\text{t}(R) = \bar{\Sigma} (R)  - \Sigma (R) \, , 
\end{equation}
where $R$ is the transverse proper distance from the assumed lens centre, $\bar{\Sigma} (R)$ is the mean density within $R$, and
\begin{equation}
\Delta \Sigma_{g{\rm t}}(R) = \frac{\Delta \Sigma_\text{t}(R)}{1  - \Sigma_\text{cr}^{-1}\Sigma (R)}\ ;
\end{equation}
the critical density for lensing, $\Sigma_\text{cr}$, can be expressed as
\begin{equation}  
\label{eq_Delta_Sigma_2}
\Sigma_\text{cr} \equiv \frac{c^2}{4\pi G} \frac{D_\text{s}}{D_\text{l} D_\text{ls}}, 
\end{equation}
where $c$ is the speed of light in vacuum, $G$ is the gravitational constant, and $D_\text{l}$, $D_\text{s}$, and $D_\text{ls}$ are the angular diameter distances to the lens, to the source, and from the lens to the source, respectively.

For a source redshift distribution, the source averaged reduced excess surface mass density can be approximated as \citep{ume20},
\begin{equation}
\label{eq_esd_2}
\langle \Delta \Sigma_{g\text{t}}  \rangle \simeq \frac{ \Delta \Sigma_\text{t}}{1- \langle \Sigma_\text{cr}^{-1}  \rangle \Sigma} \;.
\end{equation}

The equations presented here and in the following hold for non-axially symmetric lenses too if azimuthally averaged quantities are considered.

\subsection{Signal measurement}

I measure the reduced excess surface density $\Delta\Sigma_{g{\rm t}}$ in circular annuli with the estimator,
\begin{equation}
\label{eq_signal_2}
\Delta \Sigma_{g{\rm t}} (R)  =  \frac{\sum_i  w_{\Delta \Sigma, i}  e_{\text{t},i} \Sigma_{\text{cr},i}} {\sum_i w_{\Delta \Sigma, i} },
\end{equation}
where
\begin{equation}
\label{eq_signal_3}
w_{\Delta \Sigma, i}  =  \Sigma_{\text{cr},i}^{-2} w_i \; .
\end{equation}
The sum runs over the sources inside each annulus centred at $R$ and $e_{\text{t},i}$ is the tangential component of the ellipticity of the $i$-th source galaxy, $w_i$ is the weight assigned to the source ellipticity, and $\Sigma_{\text{cr},i}$ is the critical density for the $i$-th source.

The shape catalogue used in this work lists measurements of the distortion $\delta$ which is related to the ellipticity through the responsivity ${\cal R}$ as \citep{man+al08,row+al15,kai+al95}, 
\begin{equation}
e_i = \frac{\delta_i}{2 {\cal R}} \, .   
\end{equation}
The mean responsivity in an annulus can be calculated from the inverse variance weights and the per-object estimates of the RMS distortion $\delta_{\text{RMS},i}$ as \citep{hsc_man+al18},
\begin{equation}
\label{eq_signal_5}
\langle {\cal R} \rangle \simeq 1 - \langle \delta^2_{\text{RMS},i}  \rangle_{\Delta \Sigma} \;,
\end{equation}
where $\langle ... \rangle_{\Delta \Sigma}$ denotes an average with the $w_{\Delta \Sigma, i}$  weights.

I compute distances to the sources and critical surface densities or select background galaxies based on the photo-$z$ point-estimator. Other methods exploit the per-source photo-$z$ probability density function \citep{she+al04}, or the ensemble source redshift distribution \citep{kids_hil+al20}. These methods rely on unbiased and accurate redshift probability distribution, which need expensive calibration samples or simulations \citep{hsc_tan+al18,kids_hil+al20}. For this analysis, I prefer to rely on conservative selections based on the photo-$z$ point-estimator \citep{ser+al17_psz2lens,hsc_med+al18b}, whose level of systematic errors is under control and subdominant for Stage-III surveys \citep{bel+al19,euclid_pre_XLII_ser+al24}. 

\subsection{Background source selection}
\label{sec_back_sel}

I select galaxies behind the lens as sources for the WL analysis based on their photo-$z$ or colours. As a first step, I select galaxies such that
\begin{equation}
\label{eq_zphot_1}
z_\text{p} > z_\text{lens} +\Delta  z_\text{lens}\;,
\end{equation}
where $z_\text{lens}$ is the lens redshift, $z_\text{p}$ is the source redshift, and $\Delta  z_\text{lens}=0.1$ is a threshold above the cluster redshift to lower the contamination. 

On top of this criterion, I require that the sources pass more restrictive cuts in either photo-$z$ or colour properties, which I discuss below.

\subsubsection{Photometric redshifts}
\label{sec_back_sel_photoz}

A secure population of background galaxies can be selected with criteria based on the photo-$z$s \citep{ser+al17_psz2lens}:
\begin{eqnarray}
z_\text{p,2.3\,\%} 		 &>&	 	z_\text{lens} + \Delta  z_\text{lens} \,; \label{eq_zphot_2}  \\
0.2		&<&		z_\text{p}\,; \label{eq_zphot_3}  \\
1.5	&>&		z_\text{p}\,;   \label{eq_zphot_4} \\
R(z_\text{p})		&>&		0.8 \,,\label{eq_zphot_5} 
\end{eqnarray}
where $z_\text{p,2.3\,\%}$ is the lower bound of the region including the 95.4\,\% of the probability density distribution. The photo-$z$ risk parameter $R(z_\text{p})$ represents the expected loss for a given choice of $z_\text{p}$ as the point estimate and it quantifies the confidence in the point estimate of the redshift \citep{hsc_tan+al18}. Thresholds are conservative, and I select galaxies in the redshift range $0.2 < z_\text{p} < 1.5$ wherein photo-$z$'s are thought to be reliable \citep{hsc_tan+al18}. The contamination level is expected to be at the percent level \citep{ogu+al12,ser+al17_psz2lens,hsc_med+al18b,ume+al20}.

\subsubsection{Colour–colour}

Selection of galaxies in colour-colour space can provide complete and pure background samples and can complement the photo-$z$ selection \citep{euclid_pre_les+al24}. Here, I adopt the cuts in the $g - i$ vs. $r -z$ colour-colour space \citep{hsc_med+al18b}, which are optimised for the HSC-SSP. Blue or red galaxies are efficiently separated in this space for lenses up to $z_\text{lens}\lesssim 1.1$. The level of contamination is within the percent level \citep{hsc_med+al18b,euclid_pre_les+al24}.


\subsection{Shear calibration}
\label{sec_shea_cali}

The raw shape components of the source galaxies, $e_{\text{raw}, 1}$ and $e_{\text{raw}, 2}$, can be affected by  a multiplicative ($m$) or an additive ($c$) bias,
\begin{equation}
\label{eq_calibration_1}
e_{i} = \frac{e_{\text{raw}, i} - c_i}{1 + m}  \, \hspace{1cm} (i=1,2) \, .
\end{equation}
HSC-SSP calibrated the bias with simulations \citep{hsc_man+al18}. I correct each galaxy shape for the additive bias, whereas the multiplicative bias $m$ is averaged in each annulus \citep{hey+al12,mil+al13,vio+al15},
\begin{equation}
\label{eq_Delta_Sigma_5}
\langle m \rangle = \langle  m_i \rangle_{\Delta \Sigma} \;.
\end{equation}

\subsection{Rescaling}
\label{sec_resc}

When comparing lensing profiles, I rescale lengths by a proxy for the over-density radius
\begin{equation}
\label{eq_units_1}
r_{\Delta \hat{N}_\text{mem}} = \left(  \frac{  \hat{N}_\text{mem}  } { \hat{N}_\text{mem,ref} } \right)^{1/3} \left( \frac{ H(z) } { H (z_\text{ref})}  \right)^{-2/3}   \text{Mpc} \; ,
\end{equation}
where the pivots are fixed to $\hat{N}_\text{mem,ref} = 30$ and $z_\text{ref} = 0.6$.

For each cluster, I measure the lensing signal $\Delta\Sigma_{g\text{t}}$ in eight radial circular annuli equally separated in logarithmic space spanning the range between $R_\text{min} / r_{\Delta \hat{N}_\text{mem}} = 0.3\, h^{-1}$ ($\sim0.43$) and $R_\text{max} / r_{\Delta \hat{N}_\text{mem}}=3.0\, h^{-1}$ $(\sim 4.3)$ from the cluster centre. 

Surface densities are scaled by
\begin{equation}
\label{eq_units_2}
\Sigma_{\Delta \hat{N}_\text{mem}} = 10  \left( \frac{ r_{\Delta \hat{N}_\text{mem}}}{ 1~\text{Mpc} }  \right) \left( \frac{ \rho_\text{cr}(z) } { \rho_\text{cr}(z_\text{ref})} \right)  ~M_\odot \text{pc}^{-2} \; .
\end{equation}


\subsection{Noise}
\label{sec_nois}

The uncertainty covariance matrix for a single lens accounts for measurement uncertainties and lensing from the uncorrelated large-scale structure (LSS), $\Delta \Sigma_\text{LSS}$,
\begin{equation}
\label{eq_inf_2}
\texttt{C}= \texttt{C}^\text{stat}+ \texttt{C}^\text{LSS} \;,
\end{equation}
where $\texttt{C}^\text{stat}$  accounts for the shape noise and shear measurements uncertainties and $\texttt{C}^\text{LSS}$ is the covariance due to LSS. The main source of noise is the intrinsic ellipticity dispersion $\sigma_{e_\alpha}$. For Stage-III surveys and optical bands, $\sigma_{e_\alpha} \simeq 0.26$-$0.3$ \citep{sch+al18,euc_pre_mar+al19,euclid_pre_aja+al23}.

The covariance $\texttt{C}^\text{LSS}$ accounts for lensing from LSS  \citep{hoe03,ser+al18_psz2lens}. It can be written as the cross-correlation between two angular bins $\Delta \theta_i$ and $\Delta \theta_j$ \citep{sch+al98b,hoe03},
\begin{eqnarray} 
\texttt{C}^\text{LSS}_{ij}  & = & \langle \Delta \Sigma_\text{LSS}(\Delta \theta_i)  \Delta \Sigma_\text{LSS}(\Delta \theta_j) \rangle \\
& = &  2 \pi  \Sigma_\text{cr}^2 \int_0^{\infty} P_k(l)g(l, \Delta \theta_i) g(l, \Delta \theta_j) \ l \ dl \, , \label{eq_lss_1}
\end{eqnarray}
where $P_k(l)$ is the effective lensing projected power spectrum. I compute the linear matter power spectrum with a semi-analytical fitting function \citep{ei+hu98}, and the effects of non-linear evolution with the revised halofit model \citep{tak+al12}. These approximations are adequate for the precision needed in our analysis. The function $g$ is the filter. In an angular bin $\theta_1 < \Delta \theta< \theta_2$,
\begin{equation} 
g=\frac{1}{\pi(\theta_1^2 -\theta_2^2)l} \left[ \frac{2}{l} \left( J_0(l \theta_2)  -J_0(l \theta_1) \right) +\theta_2 J_1(l \theta_2) -\theta_1 J_1(l \theta_1) \right].
\end{equation}

Here, I do not consider correlated noise due to intrinsic alignment of sources, which can be neglected for Stage-III analyses of cluster lensing \citep{mcc+al19,ume+al20}, or correlated large-scale structure, which is subdominant in the radial range I consider \citep{mcc+al19,euclid_pre_XLII_ser+al24}. 

For our reference analysis, I assume that lensing signals between lenses are uncorrelated. Our sample consists of 400 clusters in an area of $\lesssim 140\deg^2$, with an average projected separation in the plane of the sky of $\sim0.6\deg$. This is larger than the radial extension for the shear profiles under consideration, which cover $\lesssim 0.2\deg$ ($\lesssim 0.4\deg$) at the median (minimum) redshift of the sample.


\subsection{Signal-to-noise}

The signal-to-noise S/N of the WL profile can be defined in terms of the weighted excess surface density $\Delta \Sigma_{g{\rm t}}$ in the relevant radial range $ R_\text{min} <R< R_\text{max}$ \citep{ser+al17_psz2lens},
\begin{equation}
\label{eq_SNR_1}
(\text{S/N})_\text{WL}= \frac{\Delta \Sigma_{g{\rm t}} ( R_\text{min} <R< R_\text{max})}{\delta_\text{t}} \;,
\end{equation}
where the noise $\delta_\text{t}$ includes statistical uncertainties and cosmic noise. I measure $(\text{S/N})_\text{WL}$ between $R_\text{min} = 0.3\, h^{-1}~\text{Mpc}$ and $R_\text{max} =3.0\, h^{-1}~\text{Mpc}$ from the cluster centre.


\subsection{Population averages and covariances}
\label{sec_cova}

Here I describe how the average lensing profiles are computed and the related covariance matrix. The covariance can be computed theoretically, analytically, or by resampling. For this analysis, I consider the latter two methods.

\subsubsection{Analytical}

The weighted average of the lensing profiles of a population of lenses in a redshift bin, $\langle \vec{\Delta \Sigma}_{g{\rm t}} \rangle_\texttt{W}$, can be measured with standard statistical methods as \citep{sch95}
\begin{equation}
\langle \vec{\Delta \Sigma}_{g{\rm t}} \rangle_\texttt{W} = \sum_j \texttt{W}_j \cdot \vec{\Delta \Sigma}_{g{\rm t}, j} \, ,
\end{equation}
with weight matrices $\texttt{W}_j$,
\begin{equation}
\texttt{W}_j =  \left (\sum_i \texttt{C}_{ii}^{-1} \right)^{-1}  \cdot \texttt{C}_{jj}^{-1} \, ,
\end{equation}
where $\texttt{C}_{jj}$ is the covariance matrix of the $j$-th lensing profile $\vec{\Delta \Sigma}_{g{\rm t}, j}$. For the present analysis, $\texttt{C}_{ii}$  is non-diagonal to to LSS noise.

The weighted covariance matrix for the average can be written as
\begin{equation}
\texttt{C}_\texttt{W}  = \sum_{i,j} \texttt{W}_{i} \cdot  \texttt{C}_{ij} \cdot \texttt{W}_{j}^\text{T}  \, ,
\end{equation}
where $\texttt{C}_{ij}$ is the cross covariance between the $i$-th and the $j$-th lens.  For the reference analysis, the cross-covariance between different lenses ($i \neq j$) is assumed to be null.

Together with the measurements uncertainties, scatter due to rescaling should be accounted for too. The richness is a scattered proxy of the mass. If the mass is overestimated, the profile will be rescaled to a lower value of $R /r_\Delta$ than expected, and, at the same time, the value of $\Delta \Sigma_{g\text{t}} / \Sigma_\Delta$ will be biased low. The scatter in the proxy over-density radius, $r_{\Delta | \hat{N}_\text{mem} }$, can be written as
\begin{equation}
\delta_{r_\Delta | \hat{N}_\text{mem} } \sim \frac{1}{3}  
\frac {  \sigma_{ \ln \hat{N}_\text{mem} | \ln M_{\Delta\text{c}} }  }  { \sqrt{N_\text{av}} } \, ,
\end{equation}
where $N_\text{av}$ is the number of averaged clusters and  $\sigma_{ \ln \hat{N}_\text{mem} | \ln M_{\Delta\text{c}} } \sim 0.15$ is the proxy scatter \citep{ogu+al18}. The scatter in the rescaling unit $r_{\Delta | \hat{N}_\text{mem} }$ is a source of scatter for the rescaled profiles. If the lensing profile has logarithmic slope equal to $\gamma_R$, $\Delta \Sigma_\text{t} \propto R^{-\gamma_R}$, the associated covariance matrix can be written as
\begin{equation}
\texttt{C}_{\delta} = (1 + \gamma_R)^2 \delta_{r_\Delta | \hat{N}_\text{mem}}^2  \langle \vec{\Delta \Sigma}_{g{\rm t}} \rangle_{\texttt{W}}^\text{T} \cdot \langle \vec{\Delta \Sigma}_{g{\rm t}} \rangle_{\texttt{W}}  \, .
\end{equation}
$\texttt{C}_{\delta}$ is subdominant with respect to $\texttt{C}_\texttt{W}$ and a nearly isothermal profile with $ \gamma_R = 1$ can be safely assumed.

\subsubsection{Resampling}

Covariance matrices can be alternatively computed with internal estimators that resample directly the observed data using, for example, jackknife or bootstrap to generate multiple copies of the observations \citep{mo+pe22}. The added value of these approaches is that they account for all systematic or statistical uncertainties and correlations, including cross-correlations between redshift or richness bins, which are estimated with a data-driven approach. On the other hand, the inverse of a noisy, unbiased estimator of the covariance matrix is not an unbiased estimator of the inverse covariance matrix \citep{har+al07,man+al13}, and corrections or calibrations might still be needed.

For the survey under study, I consider the analytical covariance as the reference one mostly due to the small sample (only 400 clusters) and limited survey area ($\lesssim 140~\text{deg}^2$). However, for result validation, I also compute the covariance matrix by bootstrap resampling with replacement. I group the lenses in simply connected regions \citep{mcc+al19}. I consider 50 groups so that there are on average two clusters from any redshift bin per group, and resample over the lens groups $10^4$ times. I resample over the lenses, that is the lensing profiles, rather than over the sources (which would need to recompute the lensing profiles at each step) but, apart from subdominant border effects for clusters near the spatial region edges, the two different resamplings provide comparable results.


\subsection{Profile comparison}
\label{sec_comp}

Self-similarity of populations $a$ and $b$ is tested by comparing average lensing profiles,
\begin{equation}
\chi^2_{a,b} = 
\vec{\Delta}_{ab}
\cdot
\texttt{C}_{\Delta,ab} ^{-1}
\cdot
\vec{\Delta}_{ab} \, ,
\end{equation}
where the difference between the profiles can be written as
\begin{equation}
\vec{\Delta}_{ab} = \langle \vec{\Delta \Sigma}_{g{\rm t}} \rangle_{\texttt{W},a}  - \langle \vec{\Delta \Sigma}_{g{\rm t}} \rangle_{\texttt{W},b} \, ,
\end{equation}
and the total uncertainty covariance for the difference accounting for shape noise, LSS, and systematics in the rescaling is given by
\begin{equation}
\texttt{C}_{\Delta,ab} = \texttt{C}_{\texttt{W},aa} + \texttt{C}_{\texttt{W},bb} - 2  \texttt{C}_{\texttt{W},ab}  + \texttt{C}_{\delta, aa}  + \texttt{C}_{\delta, bb}   \, .
\end{equation}

The null hypothesis is that the profiles are self-similar, that is, $\vec{\Delta}_{ab}$ is consistent with a null signal. The $p$-value used as metric for comparison is computed as the probability for a $\chi^2$ distribution with $N_\text{dof}$ degrees of freedom to exceed the measured value, $p(\chi^2 > \chi_{a,b}^2; N_\text{dof})$. Here, $N_\text{dof}$ equals the number of radial bins. A very small $p$-value, that is a large $\chi^2_{a,b}$ compared to $N_\text{dof}$, means that observed profiles would be very unlikely if they were self-similar (null hypothesis).


\subsection{Abundance matching}
\label{sec_abun_matc}

The properties of a cluster sample can be estimated by matching its abundance with the expected abundance of haloes for a given cosmological model \citep{ryk+al12,ogu+al18,mur+al19}. Here, I first simulate a population of haloes in the $\Lambda$CDM framework, I then compute mock observable properties, and I finally match the simulated clusters to the observed ones.

I simulate clusters from the cosmological halo mass function,
\begin{equation}
\log M, z \sim \frac{d n}{d \log M} \frac{d V}{d z} \, ,
\end{equation}
where $n$ is the number density and $V$ the comoving volume. We model the halo mass function as in \citet{tin+al08}. In an area equal to that covered by the survey, we expect $\sim 3000$ haloes with mass larger than $10^{-13.9} M_\odot$ between $z=0.15$ and $z=1.05$ in the reference cosmology. The total number of simulated clusters is extracted from a Poisson distribution centred on the mean value. 

I select clusters according to a mass proxy $N_M$ related to the true halo mass as
\begin{equation}
\log N_M \sim {\cal N}(\log M, \sigma_{ \log \hat{N}_\text{mem} | \log M_{\Delta\text{c}} } ) \, ,
\end{equation}
where ${\cal N}$ is the normal distribution. The distribution is centred on the true value and it has the same scatter as the measured conditional scatter of the (logarithm of the) richness at a given mass,  $\sigma_{ \log \hat{N}_\text{mem} | \log M_{\Delta\text{c}} } \sim 0.063$, see Sec.~\ref{sec_clus}.

I account for uncertainty in the observed cluster redshift as
\begin{equation}
z_\text{obs} \sim {\cal N}( z, \sigma_z) \, .
\end{equation}
where $\sigma_z = 0.008\times (1 +z)$, , see Sec.~\ref{sec_clus}.
Clusters are assigned to their redshift bin based on $z_\text{obs}$.

For each redshift bin, I select the richer clusters, that is, the clusters with the larger values of $N_M$. Completeness is simulated by considering the $100 / C$ richer clusters, where $C$ is the completeness, and randomly drawing 100 of them. The CAMIRA cluster sample is highly complete for massive haloes. I assume a sample completeness of $C=0.95$.
 
I can finally study the mass distribution of the selected clusters. I perform $10^4$ simulations. 


\subsection{Halo model}
\label{sec:halo_model}

Cluster masses can be measured by fitting the lensing profiles in a fixed cosmological model \citep{ser+al17_psz2lens,ume20}. Here, I only discuss the specific setting adopted for the present analysis. The excess surface density profile of the stacked lenses can be modelled as \citep{joh+al07}
\begin{equation} 
\Delta \Sigma_\text{tot}=(1 - f_\text{mis}) \, \Delta \Sigma_\text{cen}  +  f_\text{mis} \, \Delta \Sigma_\text{mis} \, ,
\end{equation}
where $\Delta \Sigma_\text{cen}$ is the excess surface density contributed by the centred haloes,  $f_\text{mis}$ is the fraction of miscentred clusters, and $\Delta \Sigma_\text{mis}$ is the excess surface density of miscentred haloes.

The well centred haloes are described with $\Delta \Sigma_\text{cen}$. I model the lenses with Navarro, Frenk, White profiles \citep[NFW,][]{nfw96}, characterised by mass, $M_{200\text{c}}$, and concentration,  $c_{200\text{c}}$. As fitting parameters, I consider the logarithm (base 10) of mass and concentration, $\textbf{p} = (\log\!M_{200\text{c}}, \log\!c_{200\text{c}}$). Here, $\log\!M_{200\text{c}}$ is short for $\log_{10} \left[ M_{200\text{c}} / \left (10^{14} M_\odot \right) \right]$.

Cluster detection algorithms might misidentify the cluster centre or the optically detected centre might differ from the minimum of the gravitational potential. Miscentering leads to underestimate $\Delta \Sigma_\text{t}(R)$ at small scales and to bias low the measurement of the concentration. The azimuthally averaged surface density of an halo misplaced by $R_\mathrm{mis}$ from the centre of the lens plane can be computed as \citep{yan+al06}
\begin{equation} 
\Sigma(R|R_\mathrm{mis})=\frac{1}{2\pi}\int_0^{2\pi} 
\Sigma_\mathrm{cen}(\sqrt{R^2+R_\mathrm{mis}^2+2R R_\mathrm{mis}\cos
  \theta})
 \, d\theta, 
\end{equation}
where $\Sigma_\mathrm{cen}$ is the centred profile. I model the distribution of off-sets with an azimuthally symmetric Gaussian distribution \citep{joh+al07,hi+wh10},
\begin{equation}
P(R_\mathrm{mis})=\frac{R_\mathrm{mis}}{\sigma_\mathrm{mis}^2}\exp
\left[ -\frac{1}{2}\left(  \frac{R_\mathrm{mis}}{\sigma_\mathrm{mis}}\right)^2 \right], 
\end{equation} where
$\sigma_\mathrm{mis}$ is the scale length. The contribution of the off-centred haloes is then 
\begin{equation} 
\Sigma_\mathrm{mis} (R)=\int
P(R_\mathrm{mis}) \Sigma(R|R_\mathrm{mis})
\, dR_\mathrm{mis} .  
\end{equation}
For the CAMIRA clusters, the typical scale length is of the order of $\sigma_\mathrm{mis} \sim 0.4~\mathrm{Mpc}$ \citep{ogu+al18}. 


\subsection{Weak lensing mass inference}
\label{sec:mass_infer}

The lens parameters are fitted with a Bayesian analysis \citep{euclid_pre_XLII_ser+al24}. The posterior probability density function of the parameters, $\textbf{p}$, given the data, $\{{\langle \Delta\Sigma_{g{\rm t}}}\rangle \}$, is written as
\begin{equation} 
p(\textbf{p}| \{{\langle \Delta\Sigma_{g{\rm t}}} \rangle \})  \propto {\cal L}(\{\langle{\Delta\Sigma_{g{\rm t}}} \rangle\} | \textbf{p}) p_\text{prior}(\textbf{p}) \;,
\end{equation} 
where \textbf{p} is a vector including the model parameters, $\cal L$ is the likelihood, and $p_\text{prior}$ is the prior. I consider non-informative priors with $-1~\le~\log\!M_{200\text{c}}~\le~2$, $0 \le \log\!c_{200\text{c}} \le 1$, $ 0 \le f_\text{off} \le 0.5$, and $0.1 \le \sigma_\text{s} / (1~\text{Mpc}) \le 0.7$.

The likelihood is ${\cal L}\propto \exp (-\chi^2/2)$, where  $\chi^2$ is written as
\begin{equation}
\chi^2 = \sum_{i,j}  \left[ 
\langle \Delta\Sigma_{g{\rm t}}\rangle_i  
- \Delta \Sigma_{g{\rm t}} (R_i | \textbf{p}) 
\right]^\text{t} \texttt{C}_{ij}^{-1}    
\left[
\langle \Delta\Sigma_{g{\rm t}}\rangle_j 
- \Delta \Sigma_{g{\rm t}} (R_j | \textbf{p}) 
\right] \;;
\end{equation}
the sum extends over the radial annuli; $\Delta \Sigma_{g{\rm t}} (R_i | \textbf{p})$ is the halo model computed at the lensing weighted radius $R_i$ of the $i$-th bin \citep{ser+al17_psz2lens}; $\langle \Delta\Sigma_{g{\rm t}}\rangle_i$ is the measured reduced excess surface density in the $i$-th bin.


\section{Results}

Here, I study the properties of the observed clusters and how they fit in the $\Lambda$CDM framework.

\subsection{Masses}
\label{sec_mass}


\begin{figure}
\resizebox{\hsize}{!}{\includegraphics{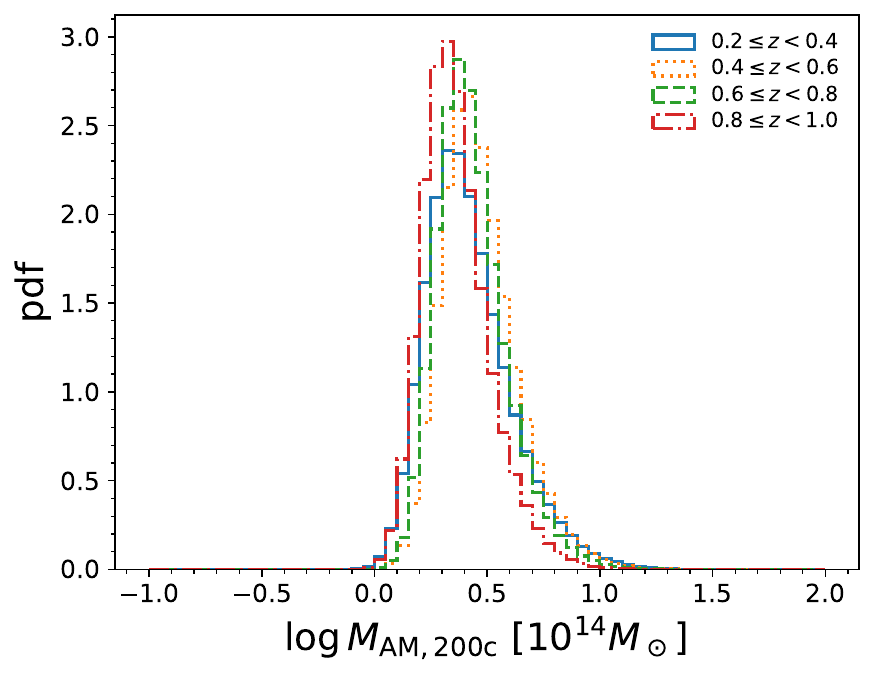}}
\caption{
Shown is the mass distribution of the selected clusters in redshift bins as estimated by abundance matching. Distributions are coded by colour and line-style according to the redshift bin.
}
\label{fig_hist_logM_AM_200c_bins}
\end{figure}


\begin{table*}
\caption{
Mass properties of the selected CAMIRA clusters for redshift bin.}
\label{tab_mass}
\centering
\begin{tabular}{@{} l r@{$\,\pm\,$}l r@{$\,\pm\,$}l r@{$\,\pm\,$}l r  r r@{$\,\pm\,$}l l l l @{}}
        \hline
        \hline
        \noalign{\smallskip}  
 	& 
\multicolumn{6}{c}{Abundance matching}  &
\multicolumn{4}{c}{Weak lensing}  &
\multicolumn{3}{c}{Mass accretion history}  \\
{} & 
\multicolumn{6}{c}{$M_\text{AM, 200c}$}  & 
$\text{S/N}_\text{WL}$ & $z_\text{WL}$ & \multicolumn{2}{c}{$M_\text{WL,200c}$} &
$M_\text{MAH,200c}$ &  \multicolumn{1}{c}{$f_\text{MAH}$}  & $z_\text{form}$ \\
$z$ range	& 
\multicolumn{2}{c}{$\text{min}$}  & \multicolumn{2}{c}{$\text{max}$} & \multicolumn{2}{c}{$\text{mean}$} & 
 \multicolumn{2}{c}{} &
 \multicolumn{2}{c}{$(z=z_\text{WL})$} & 
 \multicolumn{1}{c}{$(z=0)$} &  \multicolumn{1}{c}{$(z=z_\text{WL})$}  &  \\
 \hline
        \noalign{\smallskip} 
$[0.2,0.4)$    & 1.5 & 0.3   & 8.0 & 3.4 & 2.9  & 0.7 & 
13.6 & 0.28 & 2.4 & 0.2 &
$\sim 3.2$ & $\sim0.76$ & $\sim 0.69$ \\ 
$[0.4,0.6)$    & 1.7 & 0.3   & 8.0 & 3.0 & 3.2  & 0.7 & 
12.6 & 0.48 & 3.7 & 0.4 &
$\sim 6.2$ & $\sim0.60$ & $\sim 0.63$ \\ 
$[0.6,0.8)$    & 1.6 & 0.3   & 6.8 & 2.4 & 2.9  & 0.6 & 
8.6  & 0.68 & 2.8 & 0.4 &
$\sim 6.0$ & $\sim0.47$ & $\sim 0.64$ \\ 
$[0.8,1.0)$    & 1.4 & 0.2   & 5.6 & 1.9 & 2.5  & 0.4 & 
4.5  & 0.87 & 2.6 & 0.9 &
$\sim 7.0$ & $\sim0.37$ & $\sim 0.62$ \\ 
\hline
        \end{tabular}
\tablefoot{
Mass properties for redshift bin (col.~1) as inferred from abundance matching (cols.~2-4), WL (cols.~5-7), and mass accretion history (cols.~8-10). The redshift bin is listed in col.~1. The minimum, maximum, or mean mass as inferred from abundance matching are listed in cols.~2-4, respectively. WL determined signal-to-noise, mean redshift, and mean mass are in cols.~5-7, respectively. The extrapolated mass at $z=0$, the fraction of accreted mass at the cluster redshift, and the formation redshift, as inferred from a mass accretion history model, are reported in cols.~8-10, respectively. Masses are in units of $10^{14}M_\odot$.
}
\end{table*}



Summary statistics for the masses estimated with abundance matching, $M_\text{AM}$, for each redshift bin are listed in Table~\ref{tab_mass}. The expected mass distributions are shown in Fig.~\ref{fig_hist_logM_AM_200c_bins}. Since the richness is a nearly unbiased, time-independent proxy of the mass, a cut in richness results approximatively in a cut in mass at $M_\text{200c} \gtrsim 1.5 \times 10^{14}M_\odot$. The mass distribution is nearly time independent too, see Fig.~\ref{fig_hist_logM_AM_200c_bins}, and peaked at $M_\text{200c} \sim 3\times10^{14}M_\odot$.


\begin{figure}
\resizebox{\hsize}{!}{\includegraphics{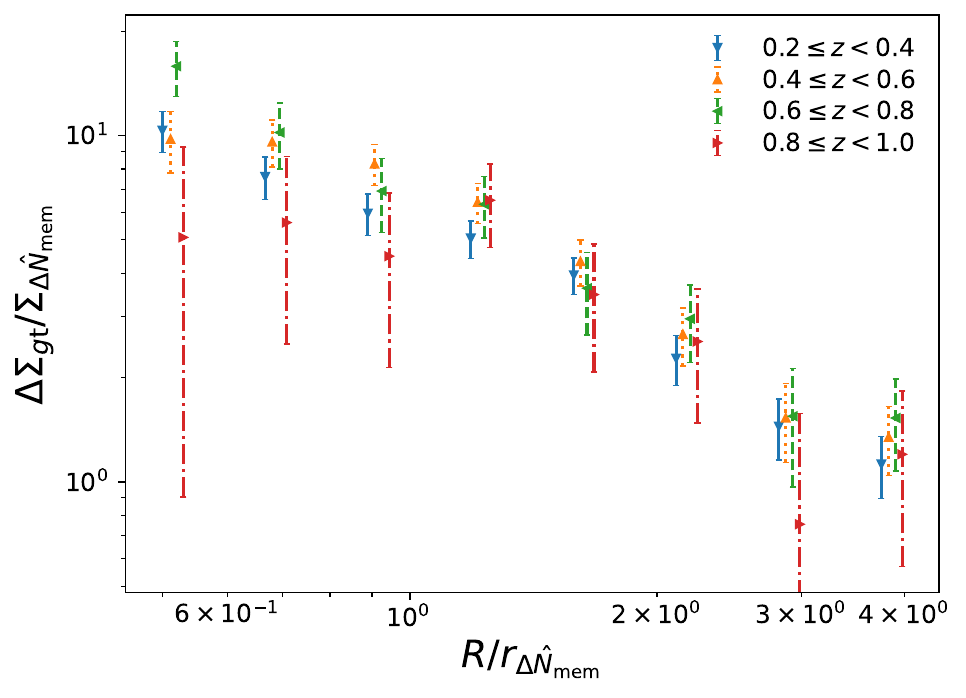}}
\caption{
Shown are the average reduced excess surface density profiles of massive CAMIRA clusters as a function of distance from the lens centre. Lengths and densities are rescaled by over-density units. 
The vertical error bars represent the square root of the diagonal elements of the total uncertainty covariance matrix, including statistical and LSS noise. Profiles are coded by colour and style according to the cluster redshifts, as in the legend, and they are horizontally shifted by 2\,\% along the abscissa to favour visualisation.
}
\label{fig_shear_profiles_delta_distance}
\end{figure}

I measured the WL signal in circular annuli.  The average excess surface density profiles are shown in Fig.~\ref{fig_shear_profiles_delta_distance}. The lensing signal is recovered with high signal-to-noise, $\text{S/N}_\text{WL}$, in each redshift bin, see Table~\ref{tab_mass}. 

The typical mass in each redshift bin can be directly inferred fitting the measured lensing  profile with a parametric lens model, see Sec.~\ref{sec:mass_infer}. The cluster mass distribution can be recovered with the WL analysis of the excess surface density $\Delta\Sigma_{g\text{t}}$. All the matter along the line of sight contributes to the lensing phenomenon. In a stacked sample of lenses, there are three main contributors: (i) the centred lenses, that is, the collapsed and nearly virialised clusters, whose signal can be well recovered up to radii $\lesssim 3~\text{Mpc}$, and whose centre is well determined, see Sec.~\ref{sec:halo_model}; (ii) clusters with the same properties of the centred ones but whose estimated centres are measured with an off-set distribution, see Sec.~\ref{sec:halo_model}; (iii) the uncorrelated matter of the LSS which fills the line-of-sight and which contributes a noise, see Sec.~\ref{sec_nois}.
The halo model is then fitted to the lensing profiles (in proper units) to give a direct estimate of the WL mass.  

The WL masses, see Table~\ref{tab_mass}, are in good agreement with estimates from abundance matching, which is a comforting check for the tested cosmological scenario. The difference between the WL mass of the mean profile and the mean mass of the abundance matched clusters is comparable to the statistical uncertainty of the WL mass and it is less than the scale on the expected distribution in mean mass from abundance matching. Due to the still limited survey area and the rarity of massive clusters, the maximum mass from abundance matching can strongly vary among different realisations of the simulated survey. On the other hand, the minimum mass, which is better sampled, is estimated with a better precision. For any deviation from $\Lambda$CDM to emerge as difference between the estimated mass from either WL or abundance matching, a larger survey area is needed.


\subsection{Self-similar density profiles}


\begin{table}
\caption{Comparison of the average reduced excess surface density profiles in different redshift bins.}
\label{tab_pvalues}
\centering
\begin{tabular}{@{} l  r r r @{}}
	\hline
	\hline
        \noalign{\smallskip} 
	$z$ range			&	$[0.4,0.6)$ &					$[0.6,0.8)$ 		&     $[0.8,1.0)$	 \\ 
	\hline
	\noalign{\smallskip}  
	$[0.2,0.4)$		&	$\chi^2_{a,b}=6.2$ ($p_{a,b}=0.63$)	&	6.4 (0.60)		&	3.6 (0.89)	\\
	$[0.4,0.6)$		&	--							&	4.2 (0.84)		&	5.4 (0.71)	 \\
	$[0.6,0.8)$		&	--							& 	 --			&	7.4 (0.49) \\
	\hline
	\end{tabular}
\tablefoot{
 For each pair of bins, I report the $\chi^2_{a,b}$ between the average profiles (with $N_\text{dof} =8$ degree of freedom), and (in round brackets) the probability to exceed this value $p_{a,b} = p\left(\chi^2 > \chi_{a,b}^2\right)$.
}
\end{table}


The mass density profile is supposed to be self-similar in a $\Lambda$CDM scenario if radial and density units are rescaled by the over-density radius and the critical density of the universe at the cluster redshift, respectively \citep{leb+al18}. The richness is a reliable proxy for the mass, and it can be used to define units for rescaling, $r_{\Delta \hat{N}_\text{mem}}$ and $\Sigma_{\Delta \hat{N}_\text{mem}}$ for length and surface density, respectively, see Sec.~\ref{sec_resc}. This can add an additional scatter, see Sec.~\ref{sec_cova}, but makes our results independent of the WL inferred mass inference, which could reduce possible time-evolution, and of the halo model.

The agreement between the profiles is good, see Fig.~\ref{fig_shear_profiles_delta_distance}, with no sign of redshift evolution. I quantify the level of agreement between signals from different redshift bins with a $p$-value statistics, see Sec.~\ref{sec_comp}. All profiles are consistent within the statistical uncertainties, with no signs for evolution, see Table~\ref{tab_pvalues}. 

Self-similarity between profiles is still recovered if the uncertainty covariance is computed with a bootstrap resampling of clusters grouped in spatially connected regions of the survey area. Notwithstanding the noisy covariance matrix, the average $p$-value is of the order of $\sim0.15$.

I excise from the analysis the cluster region within $\lesssim 0.4~\text{Mpc}$ since measured signal here could deviate from self-similarity. Firstly, miscentring can flatten the profile causing an apparent breaking of self-similarity. Secondly, in a two-phase accretion model for halo mass growth, material falling in remains in the outer regions while the core evolves almost unperturbed. Since the inner regions are the most contaminated too, I conservatively chose not to include them in WL analysis. With an even more conservative excision of the regions up to $\sim 0.57~\text{Mpc}$, the $p$-values improve on average by $\sim + 0.13$.

Since in each redshift bin the signal is dominated by clusters of similar mass, $M_\text{200c} \sim 3 \times 10 ^{14} M_\odot$, see Fig.~\ref{fig_hist_logM_AM_200c_bins}, self-similarity should be still in place if signal is stacked in comoving \citep{ume+al20} or proper units. This is in fact the case, as the average $p$-values go down by $\sim -0.15$ or  $\sim -0.24$ for comoving or proper units, respectively, but at $\sim 0.54$. or $\sim 0.46$, they are still consistent with self-similarity. At fixed mass, rescaling in comoving units captures most of the self-similarity.


\subsection{Mass accretion history}

Our sample consists of the most massive clusters in the redshift and area volume covered by HSC-SSP S16a. It does not track an evolutionary sequence. The massive CAMIRA clusters at $z \sim 1.0$ are not the progenitors of the low redshift ones at $z\sim0.2$. Their descendants are more massive but they can be missed mostly due to small volume of the observed local universe. Thanks to models for the mass accretion history of DM haloes, clusters can be placed  on an evolution track. The formation redshift $z_\text{form}$ can be defined as the redshift where the main halo progenitor has accreted one half of the expected final mass at $z=0$ \citep{mos+al19}.

\citet{cor+al15} used the extended Press-Schechter formalism to derive the halo mass accretion history from the growth rate of initial density perturbations. I use their model to compute the mass of the main progenitor along the main branch of the merger tree as a function of redshift. I take as starting point the mass and redshift estimated from WL for the average profile and put an halo with such mass and redshift on an accretion track to predict its final mass at $z=0$ or to trace it back in time and estimate its formation redshift.

Clusters in our sample are experiencing different phases of their accretion history, see Table~\ref{tab_mass}.  The clusters in the sample up to $z\lesssim 0.5$ are observed after their formation redshift, the clusters at $z\sim 0.7$ are nearly at their formation time, that is nearly halfway in their mass accretion process, whereas clusters at $z\sim 0.9$ are caught nearly $1.3~\text{Gy}$ before  $z_\text{form}$ when they have still to accrete $\sim 2/3$ of their final mass by major mergers or slow inflow. Notwithstanding the very different evolution phases, clusters in our sample show self-similar mass profiles up to $z\sim1$.

\subsection{Summary}

The WL inferred properties of the most massive optically detected clusters in HSC-SSP S16a provide a picture consistent with the standard theory of structure formation.
\begin{itemize}
\item[--] The WL mass of the richest optically selected haloes matches well the prediction from $\Lambda$CDM.
\item[--] The halo mass distribution, as inferred from the excess surface density profiles, is remarkably self-similar up to $z\sim1$.
\item[--] Substantial self-similarity is attained in the early stages of the formation process, before the main progenitor has still to accrete one half of the expected final mass.
\end{itemize}


\section{Systematics}
\label{sec_syst}

The total level of systematic uncertainty affecting the calibration of the excess surface density for HSC-SSP 16a is of the order of $\lesssim3\,\%$ \citep{mur+al19,ume+al20}. Main sources of systematic errors are due to shear calibration, photo-$z$ uncertainties, contamination due to foreground or member galaxies, and modelling effects. This level of accuracy is adequate for the precision of our profiles, which are constrained at the level of 5 -10\,\% in the redshift bins with the best $\text{(S/N)}_\text{WL}$.

The analysis here presented relies in constraints on redshift evolution more than in absolute calibration, and a constant level of systematics, if redshift independent, would not affect our analysis. The selection of background galaxies was conservative and secure in the explored redshift range. The bias, scatter, and outlier fraction for the photo-$z$s are nearly constant up to $z_\text{p}=1.5$ \citep{hsc_tan+al18}, the maximum source redshift here employed.

Shear is calibrated at the percent level \citep{hsc_man+al18}. Furthermore, the calibration of selected sources is nearly redshift independent as found by comparing the multiplicative bias for lenses at different redshifts. For clusters at $z=0.3$, $0.5$, $0.7$, and $0.9$, I find a weighted multiplicative bias of background sources of $\langle m \rangle=-0.12\pm0.9$, $-0.12\pm0.9$, $-0.12\pm0.9$, and $-0.13\pm0.9$. A time independent multiplicative bias is consistent with the data.

Contamination should be under control too. I excise from the analysis very inner regions within a minimum distance radius of $R \gtrsim 0.4~\text{Mpc}$ , which removes most of the contamination from member galaxies.

In summary, systematics should be subdominant with respect to statistical precision. Different residual effects should be uncorrelated among them and each effect should be correlated among different redshifts bins, which neither disrupt nor enhance self-similarity.


\section{Conclusions}
\label{sec_conc}

The growth of the nonbaryonic DM theory can explain both the CMB anisotropies in the liner regime or the density profiles of massive  haloes in the highly non linear regime. The list of current anomalies is intriguing but significantly shorter than the body of evidence supporting the theory \citep{pee25}. As far as DM and gravity are the main drivers of formation and evolution, the theory is very successful. Tensions mostly arise in non-linear regimes where baryonic effects become significant and can compete with gravity \citep{efs25,mad25}. The theory is seemingly paradoxical as the matter that is dark is easy to model whereas the matter that is luminous and familiar with us undergoes mechanisms difficult to grasp. But it works.

The self-similarity of galaxy clusters is a strong prediction of CDM \citep{kai86}. Simulations found that, notwithstanding the variety of dynamical states and formation histories, the density profiles of the most massive haloes in the universe, once scaled by the critical density, are remarkably similar and converge quickly to the near-universal form displayed by relaxed systems in the local universe \citep{leb+al17,mos+al19,sin+al25}.

Observation of galaxy clusters with gravitational lensing show that the mass distribution in clusters was already established when the universe was $\sim 5.7~\text{Gyr}$ old and that the evolution has been self-similar since then. Merger activity and matter infall do not disrupt the density profiles which are already stable at $z \sim 1$. Self-similarity is in place before the cluster formation time, which for the massive clusters in our sample is $z_\text{form}\sim0.6$ - $0.7$.

DM history in rich clusters differs from the hot gas, which takes some time to reach equilibrium \citep{ser+al21}. The thermalisation epoch follows the cosmic DM -- dark energy equality at $z\sim0.33$, and lies in a gentler era of structure growth.  Even though the distribution of the ICM in massive clusters has evolved self-similarly since $z\sim 1.9$ \citep{mcd+al17,ghi+al21}, the most massive clusters in the observed universe attained an advanced thermal equilibrium only $\sim~1.8~\text{Gyr}$ ago, at redshift $z =0.14\pm0.06$, when the universe was $11.7\pm0.7~\text{Gyr}$ old  \citep{ser+al21}. DM profiles were already established at  $z\gtrsim1$ (when the Universe was $\lesssim5.7~\text{Gyr}$ old) in an early phase of the cluster growth.

Self-similarity in clusters could break in the cluster cores, which evolve nearly unperturbed in a two-phase accretion process, or in the outskirts, where matter is still falling in from the cosmic web. We should be able to investigate these regimes with the final phase of Stage-III surveys when HSC-SSP will be at completion or with the first phase of Stage-IV surveys, which started with the successful launch of the {\it Euclid} satellite \citep{euclid_I_24}.


\section*{Data availability}

HSC-SSP products, including photometry, shape, and photo-$z$ catalogues, are publicly available at \url{https://hsc-release.mtk.nao.ac.jp/datasearch}. The CAMIRA catalogues are available at \url{https://github.com/oguri/cluster_catalogs}.


\section*{Code availability}
This research used the public python packages \texttt{numpy} \citep{har+al20}, \texttt{scipy} \citep{vir+al20}, \texttt{matplotlib} \citep{hun07}, \texttt{colossus} \citep{die18}, \texttt{emcee} \citep{for+al13}, \texttt{commah} \citep{cor+al15}, and \texttt{kmeans} \citep{mcc+al19}.


\begin{acknowledgements}

The author acknowledges financial contributions from contract ASI-INAF n.2017-14-H.0, contract INAF mainstream project 1.05.01.86.10, INAF Theory Grant 2023: Gravitational lensing detection of matter distribution at galaxy cluster boundaries and beyond (1.05.23.06.17), and contract Prin-MUR 2022 supported by Next Generation EU (n.20227RNLY3, The concordance cosmological model: stress-tests with galaxy clusters).

The Hyper Suprime-Cam (HSC) collaboration includes the astronomical communities of Japan and Taiwan, and Princeton University. The HSC instrumentation and software were developed by the National Astronomical Observatory of Japan (NAOJ), the Kavli Institute for the Physics and Mathematics of the Universe (Kavli IPMU), the University of Tokyo, the High Energy Accelerator Research Organization (KEK), the Academia Sinica Institute for Astronomy and Astrophysics in Taiwan (ASIAA), and Princeton University. Funding was contributed by the FIRST program from Japanese Cabinet Office, the Ministry of Education, Culture, Sports, Science and Technology (MEXT), the Japan Society for the Promotion of Science (JSPS), Japan Science and Technology Agency (JST), the Toray Science Foundation, NAOJ, Kavli IPMU, KEK, ASIAA, and Princeton University. 

Based in part on data collected at the Subaru Telescope and retrieved from the HSC data archive system, which is operated by Subaru Telescope and Astronomy Data Center at National Astronomical Observatory of Japan.

This research has made use of NASA's Astrophysics Data System (ADS) and of the NASA/IPAC Extragalactic Database (NED), which is operated by the Jet Propulsion Laboratory, California Institute of Technology, under contract with the National Aeronautics and Space Administration.

\end{acknowledgements}




\end{document}